\documentclass[12pt]{iopart}

\usepackage{iopams}  
\usepackage{amsmath}
\usepackage{graphicx}
\usepackage{booktabs}
\usepackage{subfig}
\usepackage{amsmath, amsthm, amssymb, amsfonts}
\usepackage{thmtools}
\usepackage{graphicx}
\usepackage{setspace}
\usepackage{geometry}
\usepackage{float}
\usepackage{hyperref}
\usepackage{physics}
\usepackage[utf8]{inputenc}
\usepackage[english]{babel}
\usepackage{framed}
\usepackage[dvipsnames]{xcolor}
\usepackage{tcolorbox}
\usepackage{ marvosym }
\usepackage{diagbox}
\colorlet{LightGray}{White!90!Periwinkle}
\colorlet{LightOrange}{Orange!15}
\colorlet{LightGreen}{Green!15}
\usepackage{booktabs}
\usepackage{diagbox}
\usepackage[resetlabels,labeled]{multibib}
\usepackage{enumitem}
\newcites{Math}{Math Readings}

\begin{document}

\title[On High-Dimensional Twin-Field Quantum Key Distribution]{On High-Dimensional Twin-Field Quantum Key Distribution}

\author{Ronny Mueller$^1$,  Mujtaba Zahidy$^1$, Leif Katsuo Oxenl\o we$^1$, S\o ren Forchhammer$^1$, Davide Bacco$^2$}

\address{$^1$ Department of Electrical and Photonics Engineering, Technical University of Denmark, Lyngby, Denmark}
\address{$^2$ Department of Physics and Astronomy, University of Florence, 50019 Sesto Fiorentino, Italy}
\ead{ronmu@dtu.dk}
\vspace{10pt}

\begin{abstract}
Twin-Field Quantum Key Distribution (QKD) is a QKD protocol that uses single-photon interference to perform QKD over long distances. QKD protocols that encode information using high-dimensional quantum states can benefit from increased key rates and higher noise resilience. We define the essence of Twin-Field QKD and explore its generalization to higher dimensions. Further, we show that, ultimately, the Twin-Field protocol cannot be generalized to higher dimensions in accordance with our definition.
\end{abstract}

\newcommand{\qedwhite}{\hfill \ensuremath{\Box}}
%
%
%
%

\section{Introduction}

In recent years, Quantum Key Distribution (QKD) \cite{BENNETT20147} has witnessed remarkable progress, with the Twin-Field QKD (TF-QKD) protocol standing out as a pivotal advancement in the field \cite{Lucamarini2018}. Notably, TF-QKD protocols have achieved significant breakthroughs in communication range, breaking the linear scaling bounds of channel transmittance observed in previous protocols, such as BB84 and measurement-device-independent (MDI) QKD \cite{Lo_2012}. TF-QKD introduces a square-root scaling of channel transmittance, surpassing the PLOB bound \cite{Pirandola2017} and enabling the establishment of long-distance QKD records \cite{PhysRevLett.123.100505, PhysRevLett.124.070501, Chen2021, PhysRevLett.126.250502, PhysRevLett.130.210801}. This high-performing scaling originates in the single-photon interference that is at the heart of TF-QKD, only requiring a single photon detection for a successful measurement event. Since its introduction,  TF-QKD has been the object of intensive research effort in both experimental \cite{PhysRevApplied.11.034053, PhysRevLett.123.100506, Zhou2023, Park2022} and theoretical work \cite{Grasselli_2019, PhysRevApplied.12.024061, Yin2019, Curty2019, Wang2020, Currás-Lorenzo2021}. Many variants of the original protocol have been proposed, including the notable Send-No-Send protocol \cite{PhysRevA.98.062323} that recently achieved a transmission range of more than 1000 km \cite{PhysRevLett.130.210801}. Additional degrees of freedom, i.e. polarization, have been introduced for protocols that use redundant space to improve the error rate performance \cite{PhysRevResearch.5.023069}. Nevertheless, all proposed variations of TF-QKD are still based on qubits and binary encoding.\\

\noindent Using high-dimensional quantum states for QKD has significant advantages \cite{https://doi.org/10.1002/qute.201900038}. By using an enlarged set of states compared to binary-based QKD, more than 1 bit of information can be carried by each photon. Further, the resilience to noise is increased compared to binary-based QKD, i.e. the error threshold above which no secret-key extraction can be performed 
 grows with the dimension of the protocol \cite{PhysRevA.61.062308}. These advantages have also been confirmed experimentally \cite{PhysRevA.96.022317, PhysRevApplied.14.014051, Ding2017, Islam_2019, Zahidy:22}.\\

 \noindent In this paper, we explore the possibility of generalizing the TF-QKD protocol to higher dimensions to benefit from the aforementioned advantages of high-dimensional QKD. In Sec. \ref{sec:tf_2} and Sec. \ref{sec:example}, we set the formalism and analyze a 2 and 4-dimensional example. We then reduce the TF-QKD protocol to its core by defining required assumptions on the quantum state and setup in Sec. \ref{sec:assum}. Further, we show that these assumptions lead to a contradiction and systematic error when going to any dimension higher than 2 in Sec. \ref{sec:proof}. This is followed by a discussion and conclusion in Sec{tions \ref{sec:dis} and \ref{sec:con}, respectively. In the supplements, Sec. \ref{sec:supp}, we present a conjecture in which we argue that even when allowing for systematic error, high-dimensional TF-QKD is still not feasible.

\section{Twin-Field QKD in 2 dimensions} \label{sec:tf_2}

\noindent Before analyzing the case for higher dimensions, let us first recall a 2-dimensional TF-QKD implementation.  We limit ourselves to describing only the parts relevant to this work and focus on an abstract formulation that follows the description in the supplements of \cite{Lucamarini2018}. In general, the aim of QKD is to establish some shared secret information between two remote parties, usually called Alice and Bob,  by sending quantum states. TF-QKD is a measurement-device-independent variant of QKD \cite{Lo_2012}, i.e. the measurement is performed by a third party, Charlie. To detect a potential eavesdropper, the states are prepared in mutually unbiased bases. In the case of two bases, one can use one basis to share secret information, i.e. the code basis, and another one to test for eavesdropper interference, i.e. the test basis. In TF-QKD, Alice and Bob both send a coherent pulse to Charlie and encode information in the relative phase between them. Conceptually, this can be simplified to Alice and Bob sharing a single-photon state, each party controlling a spatial mode \cite{proof_use_single}. In both cases, first-order coherence is modulated to convey information. Both parties can apply a phase to their part of the state. In the original protocol \cite{Lucamarini2018}, the overall state can be written as

\begin{equation}
\ket{\psi_{a,b}} = \frac{e^{i\gamma_a}\ket{1}_{a_1}\ket{0}_{b_1}+e^{i\gamma_b}\ket{0}_{a_1}\ket{1}_{b_1}}{\sqrt{2}} = \frac{e^{i\gamma_a}\ket{a_0}+e^{i\gamma_b}\ket{b_0}}{\sqrt{2}},
\end{equation}

\noindent where the subscripts relate to the spatial modes of the photon (it is a superposition of being either at Bob's or at Alice's), and $\gamma_i$ denotes the applied phase of each party. We restrict ourselves to only one basis, the second basis can be treated analogously. Both parties encode a single bit by choosing between two possible phases to apply. In preparation for generalization, we denote this by both parties choosing their state coefficients from a respective set $\mathcal{S}_i$ of coefficient vectors:

\begin{align}
\mathcal{S}_A &= \{\mathbf{a}_0, \mathbf{a}_1\}\\
&= \{ (e^{i\gamma_0}), (e^{i\gamma_1})\}\\
&= \{(1), (-1)\}\\
\mathcal{S}_B &= \{\mathbf{b}_0, \mathbf{b}_1\}\\
&= \{(1), (-1)\}
\end{align}

\noindent For the original protocol \cite{Lucamarini2018}, the phases are either $0$ or $\pi$. This state is measured at the middle station (Charlie) in the X basis by applying a Hadamard rotation before the measurement in the spatial modes (via a 50:50-beamsplitter before the detectors):

\begin{equation}
    \mathbf{H}  = \frac{1}{\sqrt{2}}\begin{pmatrix}
        1 & 1\\
        1 & -1
    \end{pmatrix}.
\end{equation}

\noindent Note that all tuple combinations of the two sets $\mathcal{S}_A$ and $\mathcal{S}_B$ are created. We will denote possible states by their respective indices, 

\begin{equation}
\ket{\psi_{i,j}}_2 = \frac{a_i\ket{a_0}+b_j\ket{b_0}}{\sqrt{2}}.
\end{equation}

\noindent The four possible states are then:

\begin{align}
\ket{\psi_{0,0}}_2 = \frac{1}{\sqrt{2}}\begin{pmatrix}
    1\\1
\end{pmatrix} \quad \ket{\psi_{0,1}}_2 = \frac{1}{\sqrt{2}}\begin{pmatrix}
    1\\-1
\end{pmatrix} \quad \ket{\psi_{1,0}}_2 = \frac{1}{\sqrt{2}}\begin{pmatrix}
    -1\\1
\end{pmatrix} \quad \ket{\psi_{1,1}}_2 = \frac{1}{\sqrt{2}}\begin{pmatrix}
    -1\\-1
\end{pmatrix}. 
\end{align}

\noindent Note that $\ket{\psi_{0,0}}_2$ and $\ket{\psi_{1,1}}_2$ are equal up to a global phase, as are $\ket{\psi_{0,1}}_2$ and $\ket{\psi_{1,0}}_2$. At the same time, $\ket{\psi_{0,0}}_2$ and $\ket{\psi_{1,1}}_2$ are each orthogonal to both, $\ket{\psi_{0,1}}_2$ and $\ket{\psi_{1,0}}_2$. They form two classes of states that are equal up to a global phase inside the class but orthogonal to all states outside their own class. The measurement result that Charlie shares with Alice and Bob reveals which class the state is a member of, while knowledge of the index of their own respective contribution allows each party to interfere which exact state in the class has been measured. 
This is the basis of the Twin-Field approach that allows Alice and Bob to share a secret bit. An abstract setup of the 2-dimensional TF-QKD protocol can be seen in Figure \ref{fig:setup}. \\

\begin{figure}
    \centering
    \includegraphics[width=\linewidth]{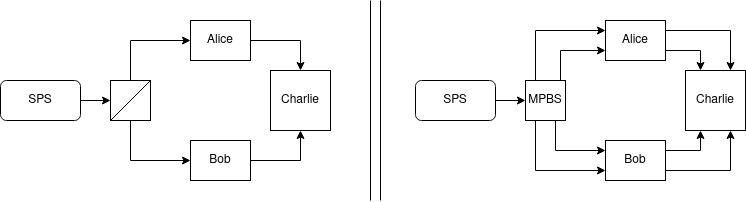}
    \caption{Possible abstract setups for a Twin-Field QKD protocol, for 2 dimensions (left) and 4 dimensions (right). SPS: Single photon source, MPBS: Multi-port beamsplitter.}
    \label{fig:setup}
\end{figure}

\section{Motivational Example of Twin-Field in 4 Dimensions}\label{sec:example}

\noindent Before going to a general analysis of TF-QKD in higher dimensions, it offers some insights to consider a concrete example first. One of the possible direct generalizations of the $\mathbf{X}$-basis measurement to 4 dimensions is realized by applying the following rotation:

\begin{equation}\label{eq:X}
    \mathbf{H}_4 = \frac{1}{\sqrt{4}}\begin{pmatrix}
        1&1&1&1\\
        1&1&-1&-1\\
        1&-1&1&-1\\
        1&-1&-1&1
    \end{pmatrix}.
\end{equation}

\noindent We extend the state to 4 dimensions as 

\begin{align}
\ket{\psi_{i,j}}_4 = &(a_{0,i}\ket{1}_{a_0}\ket{0}_{a_1}\ket{0}_{b_0}\ket{0}_{b_1}+
a_{1,i}\ket{0}_{a_0}\ket{1}_{a_1}\ket{0}_{b_0}\ket{0}_{b_1}\\
&+b_{0,j}\ket{0}_{a_0}\ket{0}_{a_1}\ket{1}_{b_0}\ket{0}_{b_1}+
b_{1,j}\ket{0}_{a_0}\ket{0}_{a_1}\ket{0}_{b_0}\ket{1}_{b_1})/2\\
=&(a_{0,i}\ket{a_0}+a_{1,i}\ket{a_1}+b_{0,j}\ket{b_0}+b_{1,j}\ket{b_1})/2,
\end{align}

\noindent and choose the sets of possible coefficient vectors as 

\begin{align}
\mathcal{S}_A &= \{\mathbf{a}_0, \mathbf{a}_1, \mathbf{a}_2, \mathbf{a}_3\}\\
&= \{(a_{0,0},a_{1,0}), (a_{0,1},a_{1,1}),(a_{0,2},a_{1,2}), (a_{0,3},a_{1,3})  \}\\
&= \{(1,1), (1,-1), (-1,1), (-1,-1)\}\\
\mathcal{S}_B &= \{\mathbf{b}_0, \mathbf{b}_1, \mathbf{b}_2, \mathbf{b}_3\}\\
&= \{(b_{0,0},b_{1,0}), (b_{0,1},b_{1,1}),(b_{0,2},b_{1,2}), (b_{0,3},b_{1,3})  \}\\
&= \{(1,1), (1,-1), (-1,1), (-1,-1)\}.\\
\end{align}

\noindent This results in 16 possible states that can be seen in Table~\ref{tab:my_table}.\\

\begin{table}[htbp]
    \centering
    \begin{tabular}{ccccc}
        \toprule
        \diagbox[height=2.5em, width=2.5em]{$i$}{$j$} & 0 & 1 & 2 & 3 \\
        \midrule
        0 & \textcolor{green}{(1,1,1,1)}& (1,1,1,-1) &(1,1,-1,-1) & \textcolor{green}{(1,1,-1,-1)}\\
        1 & (1,-1,1,1)&\textcolor{green}{(1,-1,1,-1)} &\textcolor{green}{(1,-1,-1,1)} &(1,-1,-1,-1) \\
        2 & (-1,1,1,1)&\textcolor{green}{(-1,1,1,-1)} &\textcolor{green}{(-,1,-1,1)} &(-1,1,-1,-1) \\
        3 & \textcolor{green}{(-1,-1,1,1)}&(-1,-1,1,-1) &(-1,-1,-1,1) & \textcolor{green}{(-1,-1,-1,-1)} \\
        \bottomrule
    \end{tabular}
    \caption{All possible states in our 4-dimensional example. Green states result in a deterministic measurement result when measuring using Eq. \eqref{eq:X}, whereas black states result in a uniform outcome, i.e. all measurement results are equally likely. The normalization factor $1/\sqrt{4}$ is omitted.}
    \label{tab:my_table}
\end{table}

\noindent Exactly 8 of these 16 states result in a single, conclusive measurement result, marked in green. These 8 states can be put into 4 classes similar to the case for 2 dimensions, where 2 each are equal up to a global phase but orthogonal to all other states outside their respective class. The remaining states result in a completely uniform outcome, forcing Alice and Bob to guess which values their partner chose. These ''bad" states originate from the independence in which Alice and Bob choose their coefficients which forbids them to coordinate to only create ''good" states. \\

\noindent Consider the following example from Bob's point of view. For 2 dimensions, Bob chooses $\mathbf{b}_1$. Before the measurement, the possible quantum states are $\ket{\psi_{0,1}}_2$ and $\ket{\psi_{1,1}}_2$. Charlie announces the measurement result $m$ as 0, corresponding to $(1,0)$. Given the measurement result, Bob can infer that the full state was $\ket{\psi_{1,1}}_2$, and gains 1 bit of information on $\mathbf{A}$. For 4 dimensions, we assume Bob chose $\mathbf{b}_3$, i.e. the possible states are $\ket{\psi_{0,3}}_4$, $\ket{\psi_{1,3}}_4$, $\ket{\psi_{2,3}}_4$ and $\ket{\psi_{3,3}}_4$, with uniform probabilities. The measurement result is again 0, corresponding to $(1,0,0,0)$. Using Bayes' theorem, the conditional probabilities are $[0,1/6,1/6,2/3]$ for $\ket{\psi_{0,3}}_4$, $\ket{\psi_{1,3}}_4$, $\ket{\psi_{2,3}}_4$, and $\ket{\psi_{3,3}}_4$, respectively. The resulting information gain (before considering leakage) is the difference in the entropy of the symbols prior to and after measurement, i.e. $\text{H}(\mathbf{A}) - \text{H}(\mathbf{A}_{\text{post}})= 2 - 1.25 = 0.75$. This is less than for dimension 2. Further, there is no systematic error rate for dimension 2. For dimension 4, assuming that Bob always chooses the most likely state as his guess, the systematic error rate, i.e. any error rate caused by ambiguous measurement results, is already above $33\%$. Other states follow this error rate. The values of this example can be seen in Table \ref{tab:states_meas_example}.\\

\noindent One could hope that by choosing different states and measurements one could overcome this issue. Unfortunately, we will now show that the occurrence of ambiguity is a fundamental issue that does not allow a Twin-Field style QKD to share more than one 1 bit of secret information per photon in the code basis.

\begin{table}[htbp]
    \centering
    \begin{tabular}{ccccc}

        Dimension &  Bob & Prior-measurement & Measurement & Post-measurement\\
        \toprule
        2 & $\mathbf{b}_1$ & $\ket{\psi_{0,1}}_2$,$\ket{\psi_{1,1}}_2$  & 0 $\hat{=}$ (1,0)& $\ket{\psi_{1,1}}_2$ \\
        4 & $\mathbf{b}_3$ & $\ket{\psi_{0,3}}_4$,$\ket{\psi_{1,3}}_4$,$\ket{\psi_{2,3}}_4$,$\ket{\psi_{3,3}}_4$& $0\hat{=}(1,0,0,0)$  & $\ket{\psi_{1,3}}_4$,$\ket{\psi_{2,3}}_4$,$\ket{\psi_{3,3}}_4$ \\
    \end{tabular}
    \caption{Example of information gain for 2 and 4 dimensions. Prior measurement describes all possible states the created quantum state could be, with knowledge of Bob's choice. Post-measurement describes the possible states with knowledge of Bob's choice and the measurement result shared by Charlie.}
    \label{tab:states_meas_example}
\end{table}

\section{Assumptions}\label{sec:assum}

To prove this, we first need to formalize the setting. Alice and Bob randomly choose a value that represents the secret information they want to share. To adhere to security proofs \cite{Tomamichel_2017_self, Sheridan_2010}, we assume that they choose uniformly. We denote Alice's and Bob's choice by a respective random variable, 

\begin{align}
    \textbf{A} &\sim \text{Uniform}(1,N_A)\\
    \textbf{B} &\sim \text{Uniform}(1,N_B).
\end{align}

\noindent The value of the proto-key, i.e. the $d$-ary string that is shared by Alice and Bob for later secret key extraction, is then given by an injective function of $\textbf{A}$ and $\textbf{B}$. We denote it by $\textbf{Z}$, $\mathbf{Z}=\mathbf{Z}(\mathbf{A},\mathbf{B})$. A possible example for $\mathbf{Z}$ is  $\textbf{Z} = \mathbf{A} \otimes \mathbf{B}$, with $\otimes$ being addition in a finite field.\\

\noindent The improved range-scaling of TF-QKD originates in its single-photon interaction, i.e. only a single detection is required each round. The quantum state that is measured at the middle station (Charlie) is created by two remote parties (Alice \& Bob), where each party chooses the coefficients of a fixed subset of basis states. This could for example be the different cores of a multicore fiber on which the information is sent. Our first assumption is therefore the use of a single photon state, where information is encoded in its coefficients.\\

\noindent \textbf{Assumption 1: First-Order/Single photon:} Given a $d_{\text{H}}$-dimensional Hilbert space $\mathcal{H}$ and the Fock space $\mathcal{H}_F$ we define all allowed quantum states $\ket{\Psi_{i,j}}$ to be single photon states, such that

\begin{equation}
    \ket{\Psi_{i,j}} = \ket{\psi_{i,j}}\ket{n=1}.
\end{equation}

\noindent Here, $\ket{\Psi_{i,j}}$ is on the combined Hilbert space $\mathcal{H} \otimes \mathcal{H}_F$, with $n$ labeling the photon number. We omit the Fock space for the rest of our considerations as it reduces to a single state in its subspace. $\ket{\psi_{i,j}}$ is bipartite and can be written as

\begin{equation} \label{eq:state_def}
    \ket{\psi_{i,j}} = \sum_{l=1}^{n_A} a_{l,i}\ket{a_l} + \sum_{l=1}^{n_B} b_{l,j}\ket{b_l},
\end{equation}

\noindent with $n_A+n_B = d_{\text{H}}$ and $\ket{a_l}$ and $\ket{b_l}$ forming a basis of $\mathcal{H}$.  \\

\noindent Requiring the state to be of this form is essential to TF-QKD as it results in improved range scaling compared to other (measurement-device-independent) QKD variants, i.e., only a single detection is required. Notably, we could replace the requirement of $\ket{n=1}$ with a quantum state representing a  coherent weak pulse instead, as commonly used in implementations. The deciding factor for the arguments in this work  is the encoding in the phase between the states and not the form of the states themselves. The subspaces spanned by $\ket{a_i}$ and $\ket{b_i}$ correspond to the parts of the system in control of Alice and Bob, respectively. We denote the coefficients using vectors over their subspace, i.e. 
\begin{equation}
    \mathbf{a}_i = \sum_{l=1}^{n_A} a_{l,i} \ket{a_l} \quad \text{and} \quad \mathbf{b}_i = \sum_{l=1}^{n_B} b_{l,i} \ket{b_l},
\end{equation}

\noindent such that 

\begin{equation}
    \ket{\psi_{i,j}} = \mathbf{a}_i + \mathbf{b}_j.
\end{equation}

\noindent Let $\mathcal{S}_A$ and $\mathcal{S}_B$ denote the set of coefficient vectors,

\begin{equation}
    \mathcal{S}_A := \{\mathbf{a}_0,...,\mathbf{a}_{N_A-1}\} \quad \text{and} \quad \mathcal{S}_B := \{\mathbf{b}_0,...,\mathbf{b}_{N_B-1}\}. 
\end{equation}

\noindent The set of all possible states is denoted by $\mathcal{S}_{\text{all}} = \{\ket{\psi_{i,j}}\}$, $i\in\{0,..., N_A-1\}$ and $j\in\{0,..., N_B-1\}$. We directly connect the chosen values of $\mathbf{A}$ and $\mathbf{B}$ with creating the chosen state, e.g. if $\mathbf{A}=1$ and $\mathbf{B}=3$ then the resulting quantum state that arrives at Charlie is given by $\ket{\psi_{1,3}} = \mathbf{a}_1 + \mathbf{b_3}$. There can be fewer choices than the dimension of the Hilbert state, i.e. $N_A+N_B \leq d_{\text{H}}$, allowing for the case of embedded states into a higher dimensional Hilbert space.\\

\noindent \textbf{Assumption 2: Independence:} The coefficients $\mathbf{a}_i$ and $\mathbf{b}_j$ that form the state $\ket{\psi_{i,j}}$ are randomly and independently chosen by Alice and Bob from their respective state sets, $\mathcal{S}_A$ and $\mathcal{S}_B$, i.e. $ A \perp\!\!\!\perp B$.\\

\noindent The ''independently" part is important and states that Alice and Bob are not allowed to communicate their choice. Any attempt to share information about their choice is reducing the information that is gained by the shared measurement, e.g. Bob's information gain $I$ of a measurement can be written as

\begin{equation}
    I = \text{H}(Z) = \text{H}(A) - \text{I}(A;B).
\end{equation}

\noindent The most information $I$ shared between both parties by knowing $\mathbf{Z}$, i.e. that can be sent by such a state in TF-QKD, is then bound by 

\begin{equation}\label{eq:NN}
    I \leq \log_2(\min(N_A,N_B)).
\end{equation}

\noindent Protocols usually have  $N_A =  N_B := N$ as a result. We define a QKD protocol to be high-dimensional if the average send information is more than one bit per measured quantum system in the code basis for no errors caused by the quantum channel. \\

\noindent To formulate this notion precisely, let us first consider an upper bound on the extractable secret key information $l$ per sifted symbol of the code basis \cite{Sheridan_2010, Ding2017}:

\begin{equation}\label{eq:l_in_main}
    l = \log_2(N) -  \text{H}_{\text{HD}}(q+s) - \text{H}_{\text{HD}}(e),
\end{equation}

\noindent where $\text{H}_{\text{HD}}(x) = -x\log_2(x/(N-1)) - (1-x)\log_2(1-x)$. All error rates are defined following the underlying security proof \cite{Sheridan_2010} of Eq. \eqref{eq:l_in_main}. Therefore, the phase error rate $e$ is the rate of mismatch between $Z_{\text{Alice}}$ and $Z_{\text{Bob}}$ in the test basis. The total Quantum Bit Error Rate (QBER) $q+s$ is the respective mismatch rate in the code basis, where $s$ denotes the systematic error rate, i.e. all errors that are caused by the protocol itself, and $q$ is the error contribution by the physical implementation of the protocol, e.g. depolarization, detector inefficiency, potential eavesdropper interference, and so on.  We consider a QKD system to be high-dimensional if it extracts more information per symbol than a binary system while assuming no error contribution by the physical implementation, i.e. $e=q=0$.\\

\noindent \textbf{Definition}: A possible QKD system is considered high-dimensional iff the extractable secret key per symbol is higher than 1 for a noise-free implementation, i.e. $l>1$ for $q=0,e=0$.\\

\noindent A direct consequence of the independent choice is that all possible combinations of the two sets $\mathcal{S}_A$ and $\mathcal{S}_B$ are created, i.e. $\mathcal{S}_{\text{all}} = \mathcal{S}_A \otimes \mathcal{S}_B$.\\

\noindent Now that the form of the quantum state is established, we need to define when we consider a Twin-Field implementation as failed. In the main part of this manuscript, we consider a Twin-Field implementation as failed if, from a measurement result, Alice and Bob cannot always deterministically calculate $\mathbf{Z}$, i.e. there is an unavoidable error by design, $s>0$. 
Systematic errors can occur if Alice or Bob are forced to guess between multiple possible states, e.g. see Table \ref{tab:states_meas_example}.\\


\noindent Let $m$ be a label for Charlie's measurement result. Given their own choice $\mathbf{A}(\mathbf{B})$ and the measurement result $m$, both parties need to be able to determine the same value for $\mathbf{Z}$ to avoid any errors.\\



\noindent \textbf{Assumption 3: Error-Free:} The remaining uncertainty of $\mathbf{Z}$ is 0 after a successful measurement and knowing either Bob's or Alice's choice, i.e. $\text{H}(\mathbf{Z}|\mathbf{A}\vee \mathbf{B},m)=0$, where $m$ denotes the measurement result of Charlie.\\

\noindent A detection is considered successful if the measurement  (depending on the setting) is neither ''no result" nor ''inconclusive result". The latter refers to the ''I don't know" result of an unambiguous state discrimination setup \cite{Chefles_2000}. Note that this assumption is not directly equal to saying that all states $\ket{\psi_{i,j}}$ need to be distinguishable but rather that only all states that share at least one index need to be distinguishable.\\

\noindent Assumption 3 corresponds to $s=0$ in \eqref{eq:l_in_main}. The case of $s>0$ is handled in Sec. \ref{sec:supp}.


\section{Main Theorem and Proofs}\label{sec:proof}

\noindent Now that we have defined all assumptions that we associate with TF-QKD, we will show that they lead to a contradiction.\\

\noindent \textbf{Definition/Notation:} Two states $\ket{\psi_{i,j}}$ and $\ket{\psi_{k,l}}$ are called parallel if they are equal up to a global phase, $\ket{\psi_{i,j}} = \exp{i\varphi_{lm}} \ket{\psi_{l,m}}$.\\

\noindent \textbf{Definition/Notation:} Let $\mathcal{A}$ be a collection of sets $[ \mathcal{S}_1,...,\mathcal{S}_m]$, where each set contains a number of states  $\mathcal{S}_l = \{\ket{\psi_{l_0}},...\ket{\psi_{l_{n_i}}}\}$,  $m,n_i \in \mathbb{Z}$. Given a combination of states that is created by taking exactly one state from each set $\mathcal{S}_l$,  the sets in $\mathcal{A}$ are considered to be linearly independent if all possible combinations are linearly independent.\\

\noindent \textbf{Lemma 1:} There are at most $p_{\text{max}} = \min(N_A,N_B)$ parallel states created by different coefficient combinations. Further, states cannot be parallel if they share an index.\\ 

\noindent \textbf{Proof:} Consider that there are only $N_A$ and $N_B$ different values for the coefficients $\mathbf{a}_i$ and $\mathbf{b}_j$. For two states $\ket{\psi_{i,j}}$ and $\ket{\psi_{k,l}}$ to be equal up to a global phase but not to be the same state (i.e. $i=k \land j=l$), both their coefficient parts need to be different, $i \neq k$ $\land$ $j\neq l$. Let us  assume that this would not be true, and the states $\ket{\psi_{ij}}$ and $\ket{\psi_{kj}}$ are equal up to a global phase:

\begin{align}
    \ket{\psi_{ij}} =& e^{i\varphi} \ket{\psi_{kj}}\\
    \implies& \mathbf{a}_i = e^{i\varphi} \mathbf{a}_k\\
    &  \mathbf{b}_j = e^{i\varphi} \mathbf{b}_j\\
    \implies& e^{i\varphi}=1\\
    \implies& \mathbf{a}_i = \mathbf{a}_k\\
    \text{\Lightning} & \ket{\psi_{ij}} \neq \ket{\psi_{kj}}
\end{align}

\noindent The analog can be shown for the first index. There are only $p_{\text{max}}=\min(N_A,N_B)$ ways to combine the coefficients such that there are no states with the same coefficients and therefore there are at most $p_{\text{max}}$ different states that are equal up to a global phase. \qedwhite\\

 \noindent \textbf{Lemma 2:} Assumption 3 implies every state in $\mathcal{S}_{\text{All}}$ that is not either parallel or linearly independent to all other states in $\mathcal{S}_{\text{All}}$ introduces errors into the protocol.\\ 

\noindent \textbf{Proof:} 

 \noindent Let $\ket{\psi_{i,j}}$ be a state that is neither parallel nor linearly independent to all other states in $\mathcal{S}_{\text{All}}$. WOLG, we assume Alice's point of view. According to the results of unambiguous quantum state discrimination \cite{Chefles_2000}, there is a measurement result $m$ that could be the result of $\ket{\psi_{i,j}}$ or a state $\ket{\psi_{i,l}}$, which is part of $\mathcal{S}_{\text{All}}$ according to Assumption 2. 
 Alice needs to interfere Bob's choice to calculate the value of $\mathbf{Z}$. Given the ambiguous measurement result, she knows that the state was either $\ket{\psi_{i,l}}$ or $\ket{\psi_{i,j}}$, i.e. Bob's choice was therefore either $l$ or $j$. $\mathbf{Z}$ is injective, such that $Z(i,j) \neq Z(i,l)$. Alice therefore needs to guess between the two values and can introduce an error with a nonzero probability. From Bob's point of view, simply choose a state $\ket{\psi_{l,j}}$ and proceed in analog. \qedwhite\\

 \noindent It follows directly from Lemma 2 that if $s=0$, the set of all possible states $\mathcal{S}_{\text{all}}$ necessarily is dividable into subsets that are linearly independent, where each subset may only contain states that are parallel.\\

\noindent We can now combine these two lemmas to show that high-dimensional Twin-Field-QKD, as defined by our assumptions, is not possible.\\

\noindent \textbf{Theorem 1:} If $N_A > 2$ or $N_B>2$, it is no longer possible to satisfy Assumptions 1, 2 \& 3.\\

\noindent Without loss of generality, let Alice have 3 states to choose from and Bob 2, such that $\mathcal{S}_A = \{\mathbf{a}_{i},\mathbf{a}_{j}, \mathbf{a}_{k}\}$ and $\mathcal{S}_B = \{\mathbf{b}_{m}, \mathbf{b}_{l}$\}. If Bob has 3 states instead, the labeling is swapped accordingly. For the general case of $N_A>2$ and/or $N_{B}>2$, any subset containing at least $3$ and $2$ coefficient vectors from either party, respectively, can be chosen. Following Assumption 2, all combinations exist, i.e. the following 6 states are part of the state set $\mathcal{S}_{\text{all}}$:

\begin{align}
    \ket{\psi_{i,m}}&=
    \begin{pmatrix}
        \mathbf{a}_i\\ \mathbf{b}_m
    \end{pmatrix}, 
    \ket{\psi_{i,l}}=
        \begin{pmatrix}
        \mathbf{a}_i\\ \mathbf{b}_l
    \end{pmatrix},
    \ket{\psi_{j,m}}=
        \begin{pmatrix}
        \mathbf{a}_{j}\\ \mathbf{b}_{m}
    \end{pmatrix},\\
    \ket{\psi_{k,m}}&=
        \begin{pmatrix}
        \mathbf{a}_k\\ \mathbf{b}_m
    \end{pmatrix},
    \ket{\psi_{k,l}}=
        \begin{pmatrix}
        \mathbf{a}_{k}\\ \mathbf{b}_l
    \end{pmatrix},
    \ket{\psi_{j,l}}=
        \begin{pmatrix}
        \mathbf{a}_j\\ \mathbf{b}_{l}
    \end{pmatrix}
\end{align}

\noindent  According to Lemma 2, these states need to be dividable into linearly independent subsets that contain only parallel states. According to Lemma 1, any subset of these states can have at most 2 parallel states.\\

\noindent If we were to split this set of states into linearly independent subsets that contain parallel states, we have to consider 5 possible cases. We denote the subsets visually by black rectangles, such that all states inside a rectangle are parallel and all rectangles only contain states that are linearly independent with all other states outside their rectangle, i.e. rectangles contain states that only differ in a global phase.

\begin{enumerate}[label=\textit{Case (\Roman*)}, leftmargin=*, align=left]
    \item \label{case:1} \textbf{All states are linearly independent}:
    \begin{equation}
    \begin{array}{cccccc}
    \boxed{\textcolor{black}{\ket{\psi_{i,m}}}} \perp
    \boxed{\textcolor{black}{\ket{\psi_{k,m}}}} \perp
    \boxed{\textcolor{black}{\ket{\psi_{j,m}}}} \perp
    \boxed{\textcolor{black}{\ket{\psi_{i,l}}}} \perp
    \boxed{\textcolor{black}{\ket{\psi_{k,l}}}} \perp
    \boxed{\textcolor{black}{\ket{\psi_{j,l}}}} 
    \end{array}
    \end{equation}

    This is not possible as these states are not linearly independent which can easily be shown by counterexample:

    \begin{align}
        &\ket{\psi_{i,m}}-\ket{\psi_{i,l}} - \ket{\psi_{j,m}} - \ket{\psi_{j,l}}
        = \begin{pmatrix}
            \mathbf{a}_i\\ \mathbf{b}_m
        \end{pmatrix}
        -        \begin{pmatrix}
            \mathbf{a}_i\\ \mathbf{b}_l
        \end{pmatrix}
        -        \begin{pmatrix}
            \mathbf{a}_j\\ \mathbf{b}_m
        \end{pmatrix}
        +        \begin{pmatrix}
            \mathbf{a}_j\\ \mathbf{b}_l
        \end{pmatrix}
        = 0
    \end{align}

    \item \label{case:2}  \textbf{2 arbitrary states are parallel}:

    \begin{equation}
    \begin{array}{cccccc}
    \boxed{\textcolor{black}{\ket{\psi_{i,m}}}} \perp
    \boxed{\textcolor{black}{\ket{\psi_{k,m}}}} \perp
    \boxed{\textcolor{black}{\ket{\psi_{i,l}}}} \perp
    \boxed{\textcolor{black}{\ket{\psi_{j,l}}}} \perp
    \boxed{\textcolor{black}{\ket{\psi_{k,l}} \parallel \ket{\psi_{j,m}}}} 
    \end{array}
    \end{equation}

    The same counterexample as in  \ref{case:1} holds.\\

    \item \label{case:3} \textbf{2 sets containing 2 parallel states}:

        \begin{equation}
    \begin{array}{cccccc}
    \boxed{\textcolor{black}{\ket{\psi_{i,m}} \parallel \ket{\psi_{k,l}}}} \perp
    \boxed{\textcolor{black}{\ket{\psi_{k,m}}}} \perp
    \boxed{\textcolor{black}{\ket{\psi_{j,l}}}} \perp
    \boxed{\textcolor{black}{\ket{\psi_{i,l}} \parallel \ket{\psi_{j,m}}}} 
    \end{array}
    \end{equation}

    The same counterexample as in \ref{case:1} holds.\\

    \item \label{case:5} \textbf{3 sets containing 2 parallel states}:

    \begin{equation}
    \begin{array}{cccccc}
    \boxed{\textcolor{black}{\ket{\psi_{i,m}} \parallel \ket{\psi_{k,l}}}} \perp
    \boxed{\textcolor{black}{\ket{\psi_{k,m}} \parallel \ket{\psi_{j,l}}}} \perp
    \boxed{\textcolor{black}{\ket{\psi_{i,l}} \parallel \ket{\psi_{j,m}}}} 
    \end{array}
    \end{equation}

    This case requires the states $\{\ket{\psi_{k,l}}, \ket{\psi_{j,l}}, \ket{\psi_{i,l}}\}$ to be linearly independent. This in turn requires $\mathbf{a_k}$, $\mathbf{a_j}$, and $\mathbf{a_i}$ to be linearly independent, which is not possible for 3 states in Alice's 2-dimensional subspace.

    \item \label{case:4} \textbf{2 or fewer linearly independent subsets}:\\
    This requires a subset to have 3 or more parallel states. This is impossible, as no set can contain more than 2 parallel states according to Lemma 2.

\end{enumerate}

\noindent As the 5 cases cover all possible scenarios, it is not possible to divide $\mathcal{S}_{\text{all}}$ as required by Lemma 1, thereby violating Assumption 3. It is therefore not possible to perform an error-free high-dimensional Twin-Field QKD protocol. \qedwhite\\

\section{Discussion}\label{sec:dis}

Theorem 1 shows that immediately when going from a binary TF-QKD protocol, where each party chooses between two options, to just adding a single state option ($N_A>2$ or $N_B>2$), we introduce an unavoidable error, i.e. binary TF-QKD is the only form of TF-QKD with no systematic error, $s=0$. We showed that the core ideas that allow Twin-Field to obtain its improved scaling with respect to transmission range results in a systematic error, without any assumptions on the physical implementation. Remarkably, this is not prohibitive for the test basis, i.e., the basis used to "catch" eavesdroppers. As the values of the test basis are revealed, one can simply identify all states and discard the results of all states that are not linearly independent with respect to some set of states we choose as the correct one, thereby avoiding errors. Nevertheless, this cannot be done for the actual code basis. In this case, any attempt to identify and discard a chosen set of states that, if discarded, would allow for error-free distinguishability, fundamentally violates the results of unambiguous quantum state discrimination \cite{Chefles_2000}. This can be seen by considering that discarding states corresponds to the "I don't know"/inconclusive result case of unambiguous quantum state discrimination. Notably, this implies that a reduction of the number of states in the test basis \cite{PhysRevA.97.042347} does not solve the problem, as it is the code basis alone that prohibits HD-TF-QKD. The proof presented so far only shows that HD-TF-QKD is not possible if we require $s=0$. One could argue that it might still be possible by allowing for a small error below the respective QKD threshold. We argue against this by bounding the minimum error rate in the supplements, see Sec.\ref{sec:supp}.\\

\noindent One remarkable variant of TF-QKD, the so-called Send-No-Send (SNS) variant \cite{PhysRevA.98.062323}, is not covered by these considerations, as it allows for states that use more than a single photon. While its test basis is the same as in the original protocol, a different code basis is used that does not use phases to convey information but rather the photon number itself. QKD with a middle station/Charlie requires Charlie to perform a measurement that works in both bases, i.e. Charlie cannot change his measurement dependent on the chosen basis. In the case of the SNS protocol, this is not a problem, as one can do a photon-number resolving measurement (1 photon or no photon for binary) and the test basis measurement at the same time. This is still possible when going to higher dimensions but requires the detection of more than one photon at once, thereby losing the characteristic range-scaling of TF-QKD. An MDI protocol that we consider similar to a high-dimensional SNS protocol has been proposed in a recent work \cite{Erkılıç2023}, where the symbols are encoded in the number of photons arriving at Charlie.

\section{Conclusion}\label{sec:con}

In this work, we considered a generalization of the well-known Twin-Field QKD protocol to higher dimensions that would keep its excellent range-scaling. We showed that the independence with which Alice and Bob choose their respective state modulation unavoidably leads to a systematic error. Additionally, in the supplements, Sec. \ref{sec:supp}, we propose a conjecture in which we argue that it is also not feasible to perform high-dimensional TF-QKD if one  allows for systematic error, as the resulting minimum error is above the maximum tolerable error rate for a high-dimensional QKD system. The respective section contains the details and arguments for this conjecture, including the explicit calculations for 3 and 4-dimensional systems.


\ack

The Center of Excellence SPOC (ref DNRF123), and Marco Lucamarini for helpful comments on TF-QKD.

\section*{Competing Interests}
The authors declare no competing financial or non-financial interests.

\section*{Data Availability}
All data used in this work are available from the corresponding author upon reasonable request.


\section{Supplements} \label{sec:supp}

In this section, we argue that allowing for some error does not allow us to perform high-dimensional Twin-Field QKD either. We were not yet able to prove this rigorously for all dimensions, it is therefore formulated as a conjecture.\\

\noindent \textbf{Conjecture 1:} For all values $N>2$, the minimum systematic error rate $s$ between Alice's and Bob's version of $\mathbf{Z}$ is higher than the maximum tolerable error rate required for a high-dimensional system, $s_{\text{max}}(N)$.\\

\noindent Here, $N$ refers to the number of coefficient vectors each party can choose from, see Eq. \eqref{eq:NN}. The maximum error rate $s_{\text{max}}$ can be lower-bounded using an asymptotical calculation \cite{Ding2017}. In general, the extractable secret key $l$ per symbol can be calculated using 

\begin{equation}\label{eq:l_supp}
    l = \log_2(N) -  \text{H}_{\text{HD}}(q+s) - \text{H}_{\text{HD}}(e + s_e),
\end{equation}

\noindent where $q+s$ is the Quantum Bit Error Rate, $s$ is the systematic error, $e+s_{e}$ is the phase error, and $s_e$ is the systematic phase error. $\text{H}_{\text{HD}}(x) = -x\log_2(x/(N-1)) - (1-x)\log_2(1-x)$. As mentioned in Sec. \ref{sec:dis}, the values of the test basis are revealed, and therefore there is no systematic error for the phase error calculation even for high dimensions, i.e. $s_e=0$. The maximum sustainable error rate $s_{\text{max}}$ for which the system can still be considered high-dimensional according to our definition is then given by that value $s$  for which

\begin{equation}\label{eq:l}
    \log_2(N) -1>  \text{H}_{\text{HD}}(s),
\end{equation}

\noindent i.e., for that value of $s$ for which the secret key $l$ per symbol is above 1 bit. Example values of $s_{\text{max}}$ can be seen in Table \ref{tab:s_max}. As an example, consider again the measurement and states proposed in Sec. \ref{sec:example} for a 4-dimensional system. By always assigning the most likely state to each measurement result, the minimum error rate is $33\%$, i.e. $s=0.33$. Using Eq. \eqref{eq:l_supp}, this corresponds to a secret key extraction of $l=0.56$ bits per symbol, assuming no additional errors caused by the actual physical implementation. This is less than the $l=1$ bit a binary TF-QKD system would deliver. It is also less robust to additional physical error, as just an additional $q=e=4\%$ error caused by the implementation results in $l=0$, i.e. no more secret key extraction is possible. This is compared to the well-known $q=e=11\%$ in a binary QKD system, i.e. the binary system is more resilient to noise. This is also what motivated our definition of a high-dimensional QKD system. The main advantages of a high-dimensional QKD system are its higher key rate per signal and its resilience to noise. A QKD protocol that uses qu$d$its but cannot have either of these characteristics should not be considered a (successful) high-dimensional QKD protocol.\\

\begin{table}[]
    \centering
    \begin{tabular}{c|ccccc}
        $N$ & $2$  & $3$ & $4$ & $5$ & $6$\\
        \hline
        $s_{\text{max}}$ & $0$ & $10.5\%$  & $19.5\%$  & $25.5\%$ & $30.5\%$  
    \end{tabular}
    \caption{Maximum systematic error values for different dimensions, calculated using Eq. \eqref{eq:l}.}
    \label{tab:s_max}
\end{table}

\noindent To justify the conjecture, we first consider the case of a 3-dimensional setup. Alice and Bob both have three coefficient vectors to choose from. There are three different successful measurement outcomes, each being associated with 3 possible states. States that share the same measurement outcome, i.e. are parallel, should not share any coefficient indices. If a measurement result allows for multiple possible states with a shared index, the most likely one should always be selected to minimize the error rate. The other states will never be chosen, resulting already in an $11.1\%$ error rate for each pair of states that shares the same index and has the same measurement result assigned. The only possible configuration, up to a permutation of indices, is therefore the following, where each box denotes states associated with the same measurement outcome:

    \begin{equation}
    \begin{array}{ccc}
    \boxed{\textcolor{black}{\ket{\psi_{0,0}},\ket{\psi_{1,2}},\ket{\psi_{2,1}}}} \quad
    \boxed{\textcolor{black}{\ket{\psi_{2,0}},\ket{\psi_{1,1}},\ket{\psi_{0,2}}}} \quad
    \boxed{\textcolor{black}{\ket{\psi_{1,0}},\ket{\psi_{0,1}},\ket{\psi_{2,2}}}} 
    \end{array}
    \end{equation}

\noindent We want to establish a lower bound on the error rate of deciding that the measured state was $\ket{\psi_{0,0}}$. Due to the symmetry of the problem, the error rate is the same for all states. Consider a subset of states that share the same second index, these states cannot be assigned to the same measurement outcome but rather need to be distinguished:

\begin{equation}
    \begin{array}{c}
    \boxed{\textcolor{black}{\ket{\psi_{0,0}}}} \quad
    \boxed{\textcolor{black}{\ket{\psi_{1,0}}}} \quad
    \boxed{\textcolor{black}{\ket{\psi_{2,0}}}} 
    \end{array}
    \end{equation}
    
\noindent To distinguish these three states with minimized error chance we need to minimize the cumulative overlap $\mathcal{A}$,

 \begin{equation}
     \mathcal{A} = |\braket{\psi_{0,0}}{\psi_{1,0}}|^2 + |\braket{\psi_{0,0}}{\psi_{2,0}}|^2 + |\braket{\psi_{2,0}}{\psi_{1,0}}|^2.
 \end{equation}

\noindent We assume that the photon is equally likely to be either at Bob or Alice, such that $|\mathbf{a}_i|^2 = |\mathbf{b}_i|^2$ = 1/2. In a slight abuse of notation of the inner product, we get

\begin{equation}
    \mathcal{A} = |\braket{\mathbf{a}_0}{\mathbf{a}_1} + 1/2|^2 +|\braket{\mathbf{a}_0}{\mathbf{a}_2} + 1/2|^2 + |\braket{\mathbf{a}_2}{\mathbf{a}_1} + 1/2|^2.
\end{equation}

\noindent This is minimized by arranging the three states $\mathbf{a}_0$, $\mathbf{a}_1$, and $\mathbf{a}_2$ co-planar with $120^{\circ}$ between all states inside the real plane. The three states $\ket{\psi_{0,0}}$, $\ket{\psi_{1,0}}$, and $\ket{\psi_{2,0}}$ are symmetric in the sense that each state is the results of applying a unitary operator $\mathbf{V}$ onto its cyclic predecessor, 

\begin{align}
    \ket{\psi_{1,0}} &= \mathbf{V} \ket{\psi_{0,0}}\\
    \ket{\psi_{2,0}} &= \mathbf{V} \ket{\psi_{1,0}}\\
    \ket{\psi_{0,0}} &= \mathbf{V} \ket{\psi_{2,0}},
\end{align}

\noindent where $\mathbf{V}$ is the unitary that corresponds to a rotation of $120^{\circ}$ around the normal of the plane in Alice's subspace and identity in Bob's subspace. We now calculate the minimum-error measurement \cite{Helstrom1967MinimumME} that could distinguish just these 3 states. This locally optimal measurement can only be better than or equal to any measurement that considers the whole state set. The minimum-error measurement is still an open problem to find for generic states but solutions have been found for some special cases, including symmetric pure states \cite{Ban1997}. As $\ket{\psi_{0,0}}$, $\ket{\psi_{1,0}}$, and $\ket{\psi_{2,0}}$ are such symmetric pure states, the optimal minimum-error measurement can be calculated. The operators are found as \cite{inbook}

\begin{equation}
    \mathbf{\pi}_j = B^{-1/2} \ketbra{\psi_j}{\psi_j}B^{-1/2} =: \ketbra{\mu_j}{\mu_j},
\end{equation}

\noindent where 

\begin{equation}
    B = \sum_{j=1}^3 \ketbra{\psi_j}{\psi_j}.
\end{equation}

\noindent The minimum error probability $P_E$ is then given by 

\begin{equation}
    P_E = 1 - \frac{1}{3}\sum_{j=1}^3 |\braket{\mu_j}{\psi_j}|^2,
\end{equation}

\noindent which can be evaluated to $P_E = 0.209$. Clearly, the best measurement on any subset of states can only perform better or equal to the best measurement on all states. We can therefore lower bound the error rate of the best global measurement by the mean of complementary subsets. The two remaining sets are $\{\ket{\psi_{0,0}, \ket{\psi_{0,1}}, \ket{\psi_{0,2}}}\}$ and $\{\ket{\psi_{0,0}}, \ket{\psi_{1,1}}, \ket{\psi_{2,2}}\}$. For the first set, the same value holds as it is the same set under symmetry between Alice and Bob. For the latter, the best possible error rate is 0, as one could choose them to be orthogonal. This allows us to calculate a lower bound on the best possible error rate using minimum error measurements as 

\begin{equation}
    s(N=3) \geq (2P_E + 1\cdot0)/3 = 0.139.
\end{equation}

\noindent This is already above $s_{\text{max}}(N=3)=10.5\%$. We expect the true error rate to be significantly worse (the bound requires the three states $\ket{\psi_{0,0}}$, $\ket{\psi_{1,1}}$, and $\ket{\psi_{2,2}}$ to be the states $\ket{\mu_j}$ and the minimum-error measurement to be identical for both subsets of states, both leading to a contradiction). 
 For 4 dimensions, the error can be bound by performing the analogous calculation,  
\begin{equation}
    s(N=4) \geq 2\cdot 0.5 + 2\cdot 0 = 0.25. 
\end{equation}

\noindent which again is above the threshold $s_{\text{max}}(N=4)=19.5\%$.

\section*{References}
\bibliography{main}

\end{document}